\documentclass [intlimits,twoside,a4paper]{article}

\usepackage{amsmath,amssymb}
\usepackage{wrapfig}

\usepackage[T2A]{fontenc}
\usepackage{graphicx}
\usepackage[cp1251]{inputenc}
\usepackage[eqsecnum]{cmpj}
\issue{2011}{14}{2}{23705}

\doinumber{10.5488/CMP.14.23705}

\title[Quantum mechanic tunneling]
{Quantum mechanic tunneling and efficiency of Faraday
current-generating process in porous nanostructures}
\author[B.A. Lukiyanets, D.V. Matulka, I.I. Grygorchak]%
{B.A. Lukiyanets, D.V. Matulka, I.I. Grygorchak}
\address{Lviv Polytechnic National University, 12~Bandera~Str., 79013~Lviv, Ukraine}

\authorcopyright{B.A. Lukiyanets, D.V. Matulka, I.I. Grygorchak, 2011}

\date{Received February 11, 2011, in final form May 31, 2011}

\begin{document}

\maketitle

\begin{abstract}
Thermodynamics and kinetics of lithium intercalation into
C--SiO${}_{2}$ nanocomposites are investigated. Dependencies of both
differential capacity and intercalation kinetics on the
nanocomposite size are established. The processes are analyzed in
terms of the impedance model. The obtained results are explained
based on the quantum effect of interference blockade of electron
tunneling into a nonmetallic nanoparticle. Propositions for the
new electrochemical energy storage technology are presented.
\keywords nanoobject, electron state, tunneling, intercalation
\pacs 71.20, 71.24, 73.40, 73.63
\end{abstract}

\section{Introduction}

A growing interest to nanostructures has been observed lately.
These structures possess a number of unique physical properties
that turned out to be quite promising from the viewpoint of their
practical applications in electronics. For instance,
nanotechnology may be one of the effective ways of solving an
urgent technical problem of producing a cathode material with
high specific energy. Really, application of the nanodispersed
 {FeS${}_{2}$} in an energy storage device with the lithium
anode increases the specific capacity  by about 20\% in comparison
with the coarse-grained homologue~\cite{1}, and the nanosized
 {$\alpha $-Fe${}_{2}$O${}_{3}$}${}_{ }$ possesses high
recirculated capacity 200~mA$\cdot$hour/g and good cycling in the
range of 1.5--4.0~V regarding   {Li${}^{+}$/Li }in comparison
with the macrostructured  {$\alpha $-Fe${}_{2}$O${}_{3}$}\,,
{$\alpha $-Fe${}_{3}$O${}_{4}$}\,, and
 {$\gamma $-Fe${}_{2}$O${}_{3 }$}~\cite{2}.

On the other hand, the urgency of searching for new cathode
materials is closely connected with lack of raw materials for
traditional cathode-active ones~\cite{3}. We think that production
of new cathode materials on the basis of cheap and ecologically
clean substances due to the dimensional (nano) effects is a
rational way  of making the energy storage devices. Beneficial
application thereof obviously depends on the degree of
understanding the physical processes in  nanostructures. The paper
is devoted to consideration of the above problem.

\section{Samples and experimental setup}

Let us consider a silicon dioxide  {SiO${}_{2 }$} which is a
widespread ecologically pure material. The nanosize silicon
dioxide intercalated by Li$^{+} $ belongs to the groups of
materials usable in developing the high capacitive energy storage
devices.

\begin{figure}[ht]
\centerline{\includegraphics[width=0.65\textwidth]{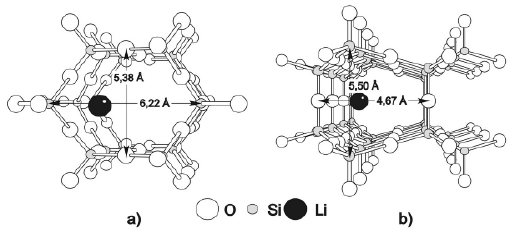}}
\caption{Structural channel in the tridymite cluster along [0001]
(parallel to C-axis) (a) and along [$11\overline{1}0]$ (normal to
C-axis) (b).} \label{fig-smp1}
\end{figure}

It is well known~\cite{4} that silicon dioxide occurs in three
structural forms, i.e., quartz, tridymite and cristobalite. None
of them in macroscale is usable as a cathode-active
material~\cite{5}. This is connected with the extremely low level
of density of ionised defects at a room temperature and low
concentration of the shallow level of trapping, although ``guest''
positions for the introduced lithium are well defined by
nanodispersed tridymite structural channels
(figure~\ref{fig-smp1}). The situation has drastically changed
with the transition to the nanosize SiO$_{2}$ with lithium
introduced into its structural channels. In this case, the density
of states at Fermi level can become considerable due to surface
states (the  ``electrochemical grafting'' concept~\cite{2,6}) and
due to sharp decrease of the diffusion resistance. Our
quantum-mechanical calculations~\cite{7} have shown that the
lithium penetration into the nanodispersed tridymite structural
channels (figure~\ref{fig-smp1}) is an exothermal process.

The nanodispersed silicon dioxide used throughout experiments was
obtained from the silicon tetrachloride by pyrogenic method in
hydrogen-air jet in the reaction  system with minimum turbulence
at 1000$\div$1500$^{\circ}$C.

\begin{wrapfigure}{o}{0.47\textwidth}
\centerline{\includegraphics[width=4cm]{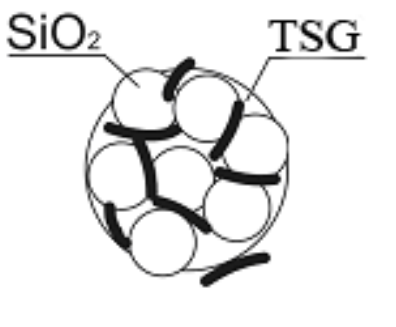}}
\caption{Fragment of the electrode structure (TSG~--
thermosplitting graphite).} \label{fig-smp2}
\end{wrapfigure}
The size of the sedimentation fractions was analyzed both by
small-angle X-ray dispersion~\cite{8} and by electron microscopy.
Electrodes with surface area 0.5~cm${}^{2}$ on a nickel substrate
were formed for electrochemical study. Composition of an electrode
is defined by ratio  {active material (SiO${}_{2}$): conductive
agent: cementing agent} as 85\% : 10\% : 5\%. The silicon dioxide
mass does not exceed 1~mg. Structure of the electrodes is
schematically shown in figure~\ref{fig-smp2}.

Thermodynamics and kinetics of the lithium intercalation were
investigated in the three-electrode electrochemical cell with the
monomolar  {LiBF${}_{4}$} solution in  {$\gamma $}-butyrolactone.
The impedance analysis was performed within the frequency range
10${}^{-2}$$\div$10${}^{5}$~Hz using AUTOLAB device manufactured
by ECO CHEMIE (Holland) equipped with FRA-2 and GPES computer
programs. Galvanostatic ``charge-discharge'' cycles were realized
by standard electronic circuit.

\begin{figure}[h]
\centerline{\includegraphics[width=10cm]{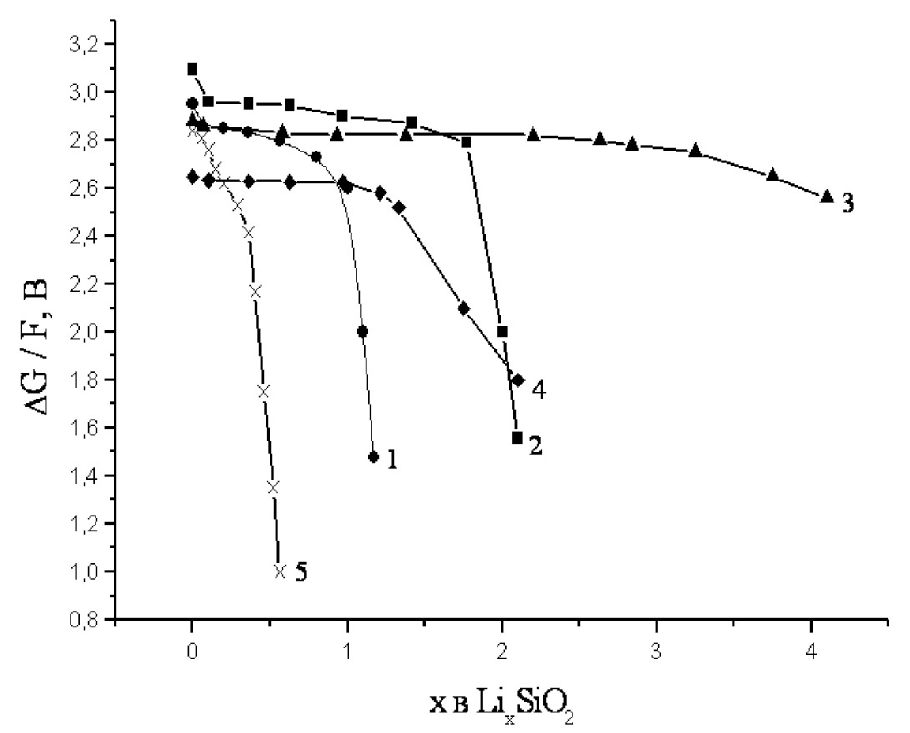}}
\caption{Gibbs function dependence on the guest lithium
concentration  in the silicon dioxide nanosystem  with average
size of particles 16~nm~(1),  11~nm~(2),   9~nm~(3), 7~nm~(4) and
5~nm~(5).} \label{fig-smp3}
\end{figure}
The electromotive force method~\cite{9} was used for thermodynamic
analysis of the process. Figure~\ref{fig-smp3} presents the
dependence of differential capacity ${\rd x}/{\rd U_{\rm rel}}$ on
$x$ ($x$ is a number lithium per formular unit of a host material;
$U_{\rm rel} $ is a relaxed value of voltage of the open circuit
relative to lithium electrode depending $x$ content). It is seen
that the current-generating reaction only for the nearly 9-nm
nanoparticles undergoes the I-st type phase transition (jump of
${\rd x}/{\rd U_{\rm rel}}$) with the formation of a two-phase
state. At $x$ defined,  X-ray investigation fixes in LiSiO${}_{2}$
the NaCl-type phase in which oxygen anions form  a face-centered
cube and the ions Li${}^{+}$ and Si${}^{4 +}$ occupy the
octahedral hollows. The lattice parameter of a such phase is
4.0287$\pm$0.0009~\AA. Figure~\ref{fig-smp3} shows a researched
current-generating reaction in the nanodispersed silicon
dioxide~\cite{7} with the average size of  the particles of the
order of 16, 11, 9 and 5~nm intercalated with lithium content of
$0.8<{x}<2$. It is seen that the specific discharge capacity,
specific energy, at $x=x_{\rm max}$ for which discharge voltage is
1.5~V, essentially depend on the nanoparticle size
(table~\ref{tbl-smp1}). The maximum flow of the reaction takes
place in case of the  9~nm size. What  exactly has caused the
situation?

\begin{table}[!h]
\caption{Specific discharge capacity.} \label{tbl-smp1}
\begin{center}\small{
\begin{tabular}{|p{0.65in}|p{0.65in}|p{0.65in}|p{0.65in}|p{0.65in}|p{0.65in}|p{0.65in}|}
\hline
 & SiO${}_{2}$ with average size of particles 16~nm & SiO${}_{2}$
  with average size of particles 11~nm & SiO${}_{2}$ with average size
  of particles 9~nm & SiO${}_{2}$ with average size of particles 5~nm &
  Macro-structured MnO${}_{2}$ \break(theory)~\cite{10} & Macro-structured
  CF${}_{x}$\break
  (theory)~\cite{10} \\ \hline
 {specific capacity, mA$\cdot$hour/g} & \vspace{2.5mm}\centerline {648} &
\vspace{2.5mm}\centerline {1080} & \vspace{2.5mm}\centerline {1890} & \vspace{2.5mm}\centerline {324}
 & \vspace{2.5mm}\centerline {310} & \vspace{2.5mm}\centerline {860} \\
\hline
\end{tabular}
}\end{center}\vspace{-0.5cm}
\end{table}
\begin{figure}[!h]
\centerline{\includegraphics[width=11cm]{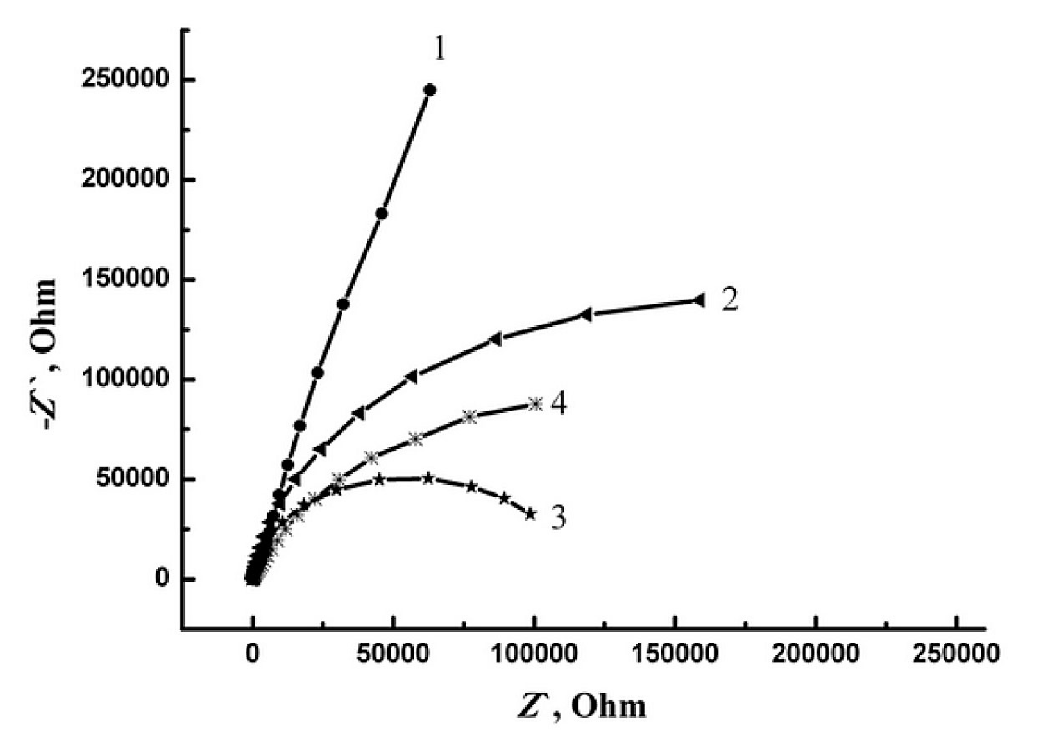}}
\caption{Nyquist diagram of the  Li$^{+}$-intercalated
current-generation for the silicon dioxide nanosystem  with
average size of particles 16~nm~(1),  11~nm~(2),   9~nm~(3), and
5~nm~(4).}\label{fig-smp4}
\vspace{-5mm}
\end{figure}

In order to analyse the Faraday's current-generating reaction in
various nanosized silicon dioxide, let us consider Nyquist
diagrams presented in figure~\ref{fig-smp4}. They show a strong
increase of the resistance  {$R{}_{\rm c}$} of the charge
transport (figure~\ref{fig-smp4}, curves~2,~4) for all sizes of
nanoparticles (with the exception of the 9-nm size in which the
kinetic mechanism takes place; curve~3) up to the practically full
blockade of the process (curve~1). In other words, the interphase
charge transport is a well-marked effect.

Herein below we present particular considerations that will enable
us to comprehend the cause and circumstances at which this
situation becomes possible.

\section{Model}

Faraday's current-generating reaction is connected with
penetration of an electron into the nanoporous hollow filled with
lithium. Today, there are some investigations of the behavior of
electrons in nanostructures. The researchers frequently use
well-known quantum mechanical models. In particular, the effect of
a resonant laser impulse irradiation on an electron within the
system of two semiconductor dots was researched in~\cite{10}. Two
interlinked rectangular wells that model the quantum dots were
used. The paper shows that quantum dynamics of electron under the
action of electromagnetic irradiation of nanostructure essentially
depends not only on the behavior of the wave but also on geometric
and energy characteristics of the dots. Resonant transition in
three-barrier nanostructure was studied in~\cite{11}. Therein a
one-dimensional model was used with three rectangular wells of
various depth, i.e., like in paper~\cite{10}, the potential on the
whole is a nonanalytic function. However, the solutions of the
problem (wave functions, energy levels) are known for each well.
The effect of rearrangement of levels in the system may be taken
into consideration from the procedure of matching the wave
functions together at the border discontinuity of the potential.

As it follows from the above cited papers, as well as from several
others, the present state of theoretical research of nanoobjects
justifies the application of even the simplest models for the
purpose of comprehending the physical effects therein.

Below we make a research within the simplest model of energy
storage in nanosystems, namely, the electron tunneling into a
nonmetallic nanoparticle in $a)$ one-dimensional model
``border-core-border'' and b) model of porous nanostructure (or
two-dimensional nanoparticle, nanotube)

\subsection{One-dimensional model ``border-core-border''}

\begin{figure}[ht]
\centerline{\includegraphics[width=6cm]{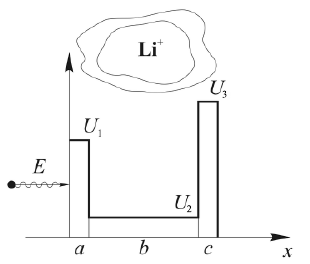}}
\caption{One-dimensional potential simulated  a nanoparticle.}
\label{fig-smp5}
\end{figure}

In  figure~\ref{fig-smp5}, within the framework of one-dimensional
problem, the nanoparticle is presented by potential
 $U_{2}$ of a hollow with two potential barriers
 $U{}_{1}$ and  $U{}_{3}$ which set off
the hollow. Different thickness of such potentials should be
regarded as~effective~potentials of particles of different sizes.
Then electron with energy  $U_{2}<{E}<\min\{ U{}_{1}\,, U_{3}\}$,
moving from left to right, will be tunnels through a barrier  {a
}into the particle. The barrier $c$ shows itself mediately, namely
through the electron de Broglie wave reflected from it.

It is well known that at certain conditions, the tunneling
probability of an electron through the system of two identical
rectangular potential barriers reaches unity~\cite{12}. This is
just the essence of resonant tunneling. This situation is observed
for a subbarrier electron with the energy value coinciding with
its virtual values in a well. This is caused by cancelation at
superposition of the tunneled electron de Broglie wave into a well
and the elastically reflected wave from the opposite barrier.
Contrary to the traditional resonant tunneling problem through the
both barriers, our interest was focused on tunneling into the
hollow, i.e., tunneling through one of them, which turns out to be
the essence of current-generating reaction.

Due to non-analyticity of the potential, it is presented as the
sum of potentials:
\begin{equation}
\label{GrindEQ__1_} U\left(x\right)=\left\{
\begin{array}{lll}
0, & x\in(-\infty, 0],&
 \left(0\right), \\
U_{1}, &  x\in\left[0,\,
x_{1} \right], & \left(1\right), \\
U_{2}, & x\in\left[x_{1}\,, x_{2} \right], & \left(2\right), \\
U_{3}, & x\in\left[x_{2}\,, x_{3} \right], &\left(3\right),\\
0, & x\in\left[x_{3}\,, \infty \right), & \left(4\right),
\end{array}
\right.
\end{equation}
i.e. all area is broken into intervals, in which the potentials
are analytical functions ($a=x_{1} -0$; $b=x_{2}-x_{1}$\,;
$c=x_{3} -x_{2}$).

The   solutions of  the time-independent Schrodinger equation are of the form:
\begin{equation} \label{GrindEQ__2_}
\psi _{s} (x)=C_{s} \exp (k_{s} x)+b_{s} \exp (-k_{s} x)
\end{equation}
with $k_{s} =\ri\sqrt{{2m\left(E-U_{s} \right)}/{\hbar ^{2} } } $;
index $s=0,{\kern 1pt} {\kern 1pt} 1,\ldots,4$ coincides with the
notations of intervals in figure~\ref{fig-smp5}.

Thus, taking into account potential~\eqref{GrindEQ__1_}, $k_{0}
=k_{4} ={\rm const}\cdot \ri\cdot \sqrt{E} $,   $k_{1} ={\rm
const}\cdot \ri\cdot \sqrt{E-U_{1} } $\,, $k_{3} ={\rm const}\cdot
\ri\cdot \sqrt{E-U_{3} } $\,, $k_{2} ={\rm const}\cdot \ri\cdot
\sqrt{E-U_{2} } $, where ${\rm const}=\sqrt{{2m}/{\hbar ^{2} } }
$\,.

In our case, tunneling through a barrier  $a$ is described by the
tunneling probability $D=\left|{C_{2} }/{C_{0} } \right|^{2}
={C_{2}^{*} C_{2} }/{C_{0}^{*} C_{0} } $\,. Thus, it is necessary
to find the weight multipliers $C_{2}$, $C_{0} $  of the
corresponding wave functions. Taking into account the properties
of the wave function (the absence of both the jump of the function
and its derivative in any point) and matching such functions and
their derivatives in the jump points  of the potential one, yields
a set of eight linear equations in $C{}_{0}$, $b{}_{0}$,
$C{}_{1}$, $b{}_{1}$, $C{}_{2}$, $b{}_{2}$, $C{}_{3}$,
$b{}_{3}$, $C{}_{4}$\, (in $\psi _{4} \left(x\right)$ { }$b_{4}
=0$ due to the absence of the reflected wave in this interval).
Analogue of such a system is shown below.

\begin{figure}[h]
\centerline{\includegraphics[width=13.7cm]{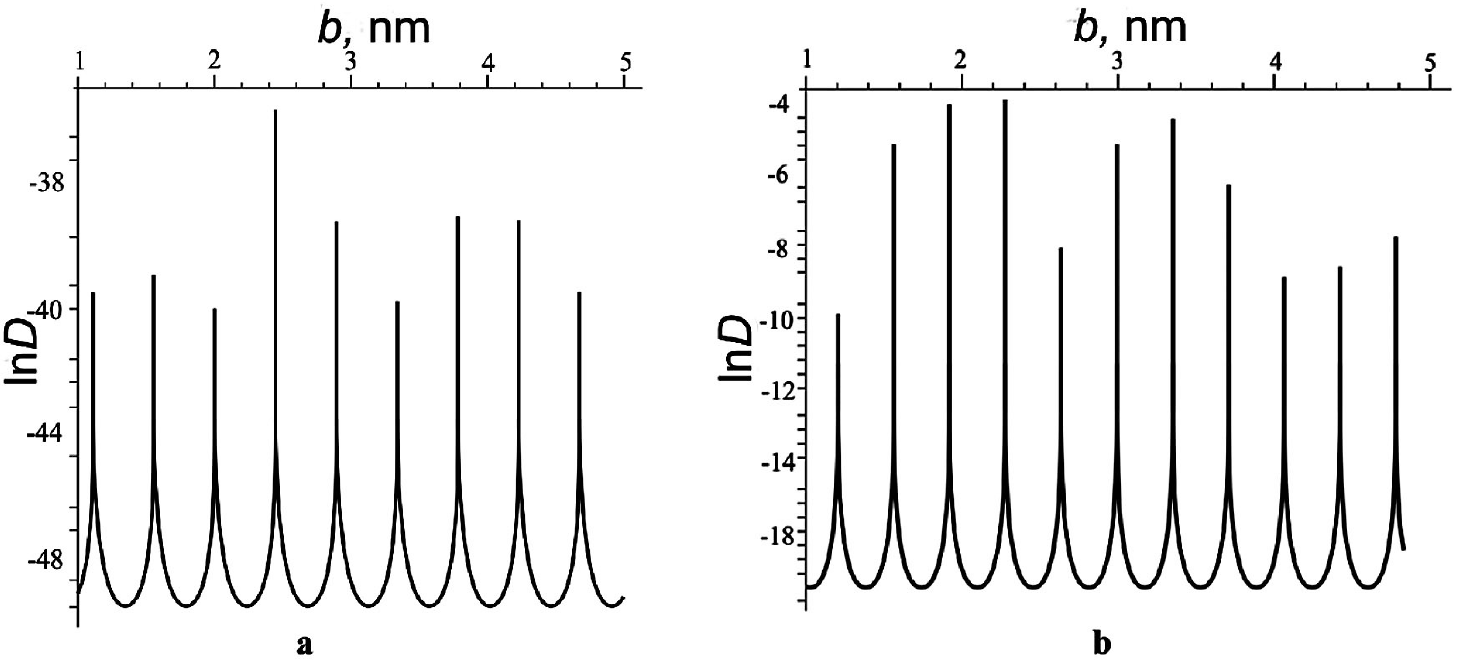}}
\caption{Dependence of the tunneling probability on the
nanoparticle size at $E=3.0$~eV~({a}) and $E=3.9$~eV~(b).}
\label{fig-smp6}
\end{figure}

A standard procedure of numerical solution of the equations was
used. In figure~\ref{fig-smp6}~(a) the logarithm of the tunneling
probability on the nanoparticle size is presented at parameters  ${a}=5$~nm, $c=4$~nm,
${U{}_{1 }}=4.0$~eV,  ${U{}_{2 }}=1.0$~eV,  ${U{}_{3 }}= 5$~eV,
${U{}_{2 }}=1.0$~eV,  ${E}=3.0$~eV, and  in figure~\ref{fig-smp6}~(b) -- at the same parameters with the only replacement of
${E}=3.0$~eV by ${E}=3.9$~eV. The replacement  also describes the
sub-barrier electron but with energy closer to the lower barrier
top.

\newpage

From figure~\ref{fig-smp6} it follows:
\begin{itemize}
\item  The property of the tunneling probability $D$ as a
function of the well thickness [$\ln D=f\left(b\right)$] is its
oscillation behavior. In the points of minimum,  $D$ is
practically equal to zero. Such behavior is a result of
superposition of the wave tunneled into  $b$ and the elastic wave
reflected from the barrier $c$.

The value of the tunneling probability essentially depends on the
ratio between the energy of electron and the height of barriers.
Comparison of figure~\ref{fig-smp6}~(a) and
figure~\ref{fig-smp6}~(b) shows that change of not only the value
but also of periodicity of the oscillation $\ln
D=f\left(b\right)$.

\item  The increased thickness  of an opposite barrier ($c$)
decreases the tunneling probability without changing its
periodicity.
\end{itemize}

Thus, depending on the ratio between geometrical and energy
characteristics of the model, the electron may or may not
penetrate into the hollow. The actual situation may be somewhat distorted
if one takes into consideration the Gaussian-like distribution of nanoparticles by size as well as energy distribution of the bombarding electrons. However this
does not change the main qualitative conclusion of the analysis,
i.e., the oscillation dependence $\ln D=f\left(b\right)$.

\subsection{Porous nanostructures (or two-dimensional nanoparticle, nanotube)}

Consider electron tunneling in porous nanostructures
(two-dimensional nanoparticle, micropores, nanotube in two models:
 {a)} with the potential
\begin{equation} \label{GrindEQ__3_3_}
U\left(x,y\right)=\left\{
\begin{array}{ll}
0\,,& x^{2}+y^{2}>R^{2}, \\
U_{1}\,, & r\leqslant\sqrt{x^{2} +y^{2}} \leqslant R, \\
U_{2}\,, & x^{2} +y^{2} <r^{2}. \end{array} \right.
\end{equation}

\begin{figure}[ht]
\centerline{\includegraphics[width=0.35\textwidth]{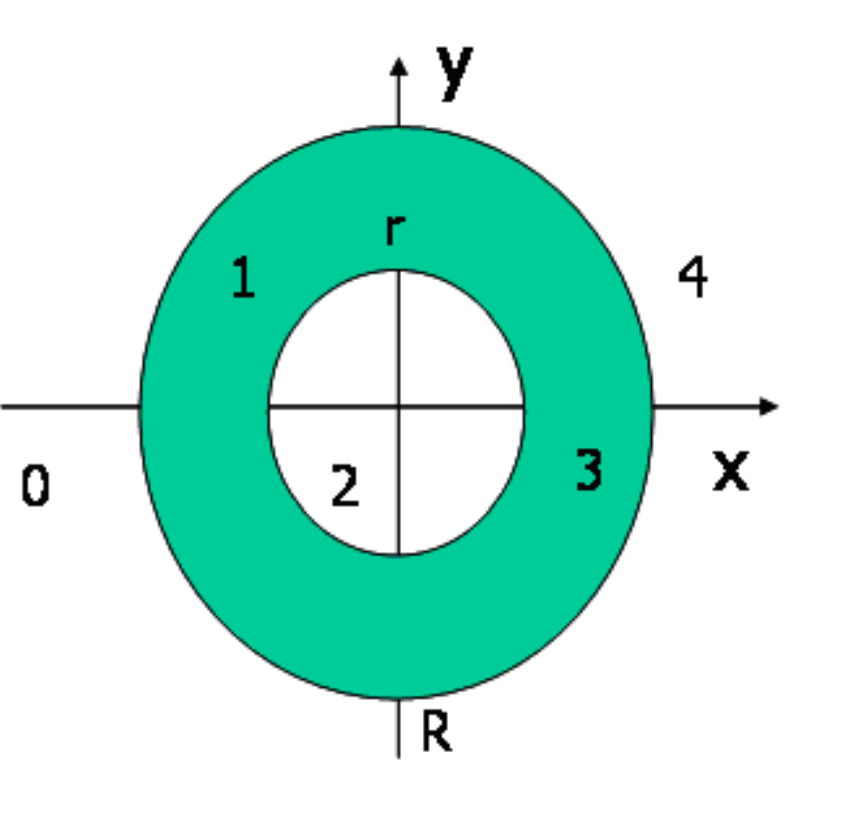}}
\caption{Two-dimensional potential simulating nanostructure
pores.} \label{fig-smp7}
\end{figure}

Here, space is divided into areas in which the potentials of the
electron are analytic functions. The model potential is presented
in figure~\ref{fig-smp7}. Therein, a pore  $U_{2} $ is restricted
by a rectangular barrier of $U_{1} $ in the form of a ring with an
outer radius $R$  and inner radius $r$ ($U_{2} <U_{1} $).

In the model  {b),} the potential is as follows:
\begin{equation} \label{GrindEQ__3_4_}
U\left(x,y\right)=\left\{
\begin{array}{ll}
0\,,&  x^{2} +y^{2} /10>R^{2},  \\
U_{1}\,, & r\leqslant \sqrt{x^{2} +y^{2} /10} \leqslant R, \\
U_{2}\,, & x^{2} +y^{2} /10<r^{2}.
\end{array}\right.
 \end{equation}

Such a model resembles the model of cylindrical tube [model
 {a)}] with the sole difference being that in this case its
cross section is an ellipse. Comparison of~\eqref{GrindEQ__3_3_}
and~\eqref{GrindEQ__3_4_} shows that the short semi-axes of
ellipses, that restricts the hollow along the  $OX$ axis,
coincides with the radii in model~{a)}. Large semi-axes of a size
$\sqrt{10}  R$ and $\sqrt{10} r$ are oriented along the $OY$ axis.

We analyze the manifestation of resonant tunneling in both models.
Herein, solutions of a time-independent Schrodinger equation are
of the form:

\begin{equation} \label{GrindEQ__3_5_}
\psi _{s} (x,y)=C_{s} \left(y\right)\exp (k_{s} x)+b_{s}
\left(y\right)\exp (-k_{s} x).
\end{equation}

Here, as well as above, $k_{s} =\ri\sqrt{{2m\left(E-U_{s}
\right)}/{\hbar ^{2} } } $, while index $s=0, 1,\ldots,4$
coincides with the notations of the ranges in
figure~\ref{fig-smp7}. Thus, $k_{0} =k_{4} ={\rm const}\cdot
\ri\cdot \sqrt{E} $, $k_{1} =k_{3} ={\rm const}\cdot \ri\cdot
\sqrt{E-U_{1} } $, $k_{2} ={\rm const}\cdot \ri\cdot
\sqrt{E-U_{2} } $, where ${\rm const}=\sqrt{{2m}/{\hbar ^{2}}}$.
The presence of $x$ and $y$ in the argument of wave function shows
the two-dimensionality of the problem.

Let the electron with energy $U_{2} <E<U_{1} $ be moving from left
to right along the $Ox$ axis. Herein, the passage through the
barrier is possible only due to tunneling. In the present case,
the barrier can be considered as a set of one-dimensional barriers
for fixed values of $y$. Geometric sizes of the barrier are varied
depending on the specific value of  {$y$}. Hence, for example, in
model~{a) }at $y=0$, the barrier is the thinnest,
$\left(R-r\right)$, while the size of the hollow is maximum,
$\left(2r\right)$. In the second limiting case, at $y=r$, two
opposite barriers interflow and their overall thickness becomes
equal to $2\sqrt{R^{2} -r^{2} } $, while the hollow thickness
becomes zero. In general, a traveling electron parallel to
 $OX$, for a fixed $y$ will be reflected from
barriers at an angle different from $\pi $. The larger is $y$ the
smaller is the angle. Below, we restrict the analysis of tunneling
in the $y$ area to no more than the neighborhood of 0.01\% of the
pore size along  $OY$ relatively to $y=0$. (As will be seen, the
$y=0$ neighborhood plays a key role in tunneling in these models).
In this case the  reflection angle practically equals $\pi$.

The tunneling probability of a homogeneous flow of electron
through the left-hand barrier into a hollow of nanoparticle is
$D\left(y\right)=\left|{C_{2} \left(y\right)}/{C_{0}
\left(y\right)} \right|^{2} ={C_{2}^{*} \left(y\right)C_{2}
\left(y\right)}/{C_{0}^{*} \left(y\right)C_{0} \left(y\right)} $,
while  the total probability is determined through the integral by
 {$y$ }from 0 up to $r$ of this characteristic. Here, $C_{0}
\left(y\right)$ is the amplitude of an incident wave, while $\,
C_{2} \left(y\right)$ is the amplitude of the wave that has
transmitted this barrier. These amplitudes may be derived from the
system of eight linear equations in respect to $C_{0}
\left(y\right)$, $b_{0} \left(y\right)$, $C_{1} \left(y\right)$,
$b\left(y\right)$, $C_{2} \left(y\right)$, $b_{2} \left(y\right)$,
$C_{3} \left(y\right)$,  $b_{3} \left(y\right)$, $C_{4}
\left(y\right)$ following from the continuity condition  of wave
functions and their derivatives in any point, including the points
of potential discontinuity. At matching together the wave
functions $\psi _{0} (x,y)$ and $\psi _{1} (x,y)$ in the point
$x_{1}(y)$, which is common for the regions 0 and 1 (it is the
root with minus sign of the equation $\sqrt{x_{1}^{2} +y^{2} }
=R$),  we shall get the following:
\begin{equation} \label{GrindEQ__3_6_}
C_{0} \left(y\right)\exp (k_{0} \cdot x_{1} )+b_{0}
\left(y\right)\exp (-k_{0} \cdot x_{1} )=C_{1} \left(y\right)\exp
(k_{1} \cdot x_{1} ) +b_{1} \left(y\right)\exp (-k_{1} \cdot x_{1}
).
\end{equation}

Similarly, matching together the wave functions
$\psi _{1} (x,y)$ and $\psi _{2} (x,y)$in the point
$x_{2} \left(y\right)$ (it is the root with minus sign
of  the equation $\sqrt{x_{2}^{2} +y^{2} } =R$)  yields the equation:
\begin{equation} \label{GrindEQ__3_7_}
C_{1} \left(y\right)\exp (k_{1} \cdot x_{2} )+b_{1} \left(y\right)
\exp (-k_{1} \cdot x_{2} )=C_{2} \left(y\right)\exp (k_{2} \cdot
x_{2} )+b_{2} \left(y\right)\exp (-k_{2} \cdot x_{2} ).
\end{equation}

Similarly, for the points $x_{2} \left(y\right)$ and $x_{3}
\left(y\right)$,  and $x_{3} \left(y\right)$ and $x_{4}
\left(y\right)$, we shall get corresponding  equations:
\begin{equation} \label{GrindEQ__3_8_}
C_{2} \left(y\right)\exp (k_{2} \cdot x_{3} )+b_{2}
\left(y\right)\exp (-k_{2} \cdot x_{3} )=C_{3} \left(y\right) \exp
(k_{3} \cdot x_{3} )+b_{3} \left(y\right)\exp (-k_{3} \cdot x_{3}
),
\end{equation}
\begin{equation} \label{GrindEQ__3_9_}
C_{3} \left(y\right)\exp (k_{3} \cdot x_{4} )+b_{3}
\left(y\right)\exp (-k_{3} \cdot x_{4} )=C_{4} \left(y\right)\exp (k_{4} \cdot x_{4} )
\end{equation}
where $x_{4} =-x_{1}$, while $x_{3} =-x_{2} $\,.

Continuity of the first derivatives in those same points yields still four equations:
\begin{equation} \label{GrindEQ__3_10_}
C_{0} \left(y\right)k_{0} \exp (k_{0} \cdot x_{1} )-b_{0}
\left(y\right)k_{0} \exp (-k_{0} \cdot x_{1} )=C_{1}
\left(y\right)k_{1} \exp (k_{1} \cdot x_{1} )-b_{1}
\left(y\right)k_{1} \exp (-k_{1} \cdot x_{1} ),
\end{equation}
\begin{equation} \label{GrindEQ__3_11_}
C_{1} \left(y\right)k_{1} \exp (k_{1} \cdot x_{2} )-b_{1}
\left(y\right)k_{1} \exp (-k_{1} \cdot x_{2} )=C_{2}
\left(y\right)k_{2} \exp (k_{2} \cdot x_{2} )-b_{2}
\left(y\right)k_{2} \exp (-k_{2} \cdot x_{2} ),
\end{equation}
\begin{equation} \label{GrindEQ__3_12_}
C_{2} \left(y\right)k_{2} \exp (k_{2} \cdot x_{3} )-b_{2}
\left(y\right)k_{2} \exp (-k_{2} \cdot x_{3} )=C_{3}
\left(y\right)k_{3} \exp (k_{3} \cdot x_{3} )-b_{3}
\left(y\right)k_{3} \exp (-k_{3} \cdot x_{3} ),
\end{equation}
\begin{equation} \label{GrindEQ__3_13_}
C_{3} \left(y\right)k_{3} \exp (k_{3} \cdot x_{4} )-b_{3}
\left(y\right)k_{3} \exp (-k_{3} \cdot x_{4} )=C_{4}
\left(y\right)k_{4} \exp (k_{4} \cdot x_{4} ).
\end{equation}

Within the system of eight homogeneous linear
equations~\eqref{GrindEQ__3_6_}--\eqref{GrindEQ__3_13_} there are
nine unknown quantities. However, having divided all the equations
into $C_{0} (y)$ we get a system of non-uniform equations with
eight unknown quantities, including $\frac{C_{2} (y)}{C_{0} (y)}
$, which is connected with the tunneling probability into the
hollow.

\begin{figure}[th]
\centerline{\includegraphics[width=0.99\textwidth]{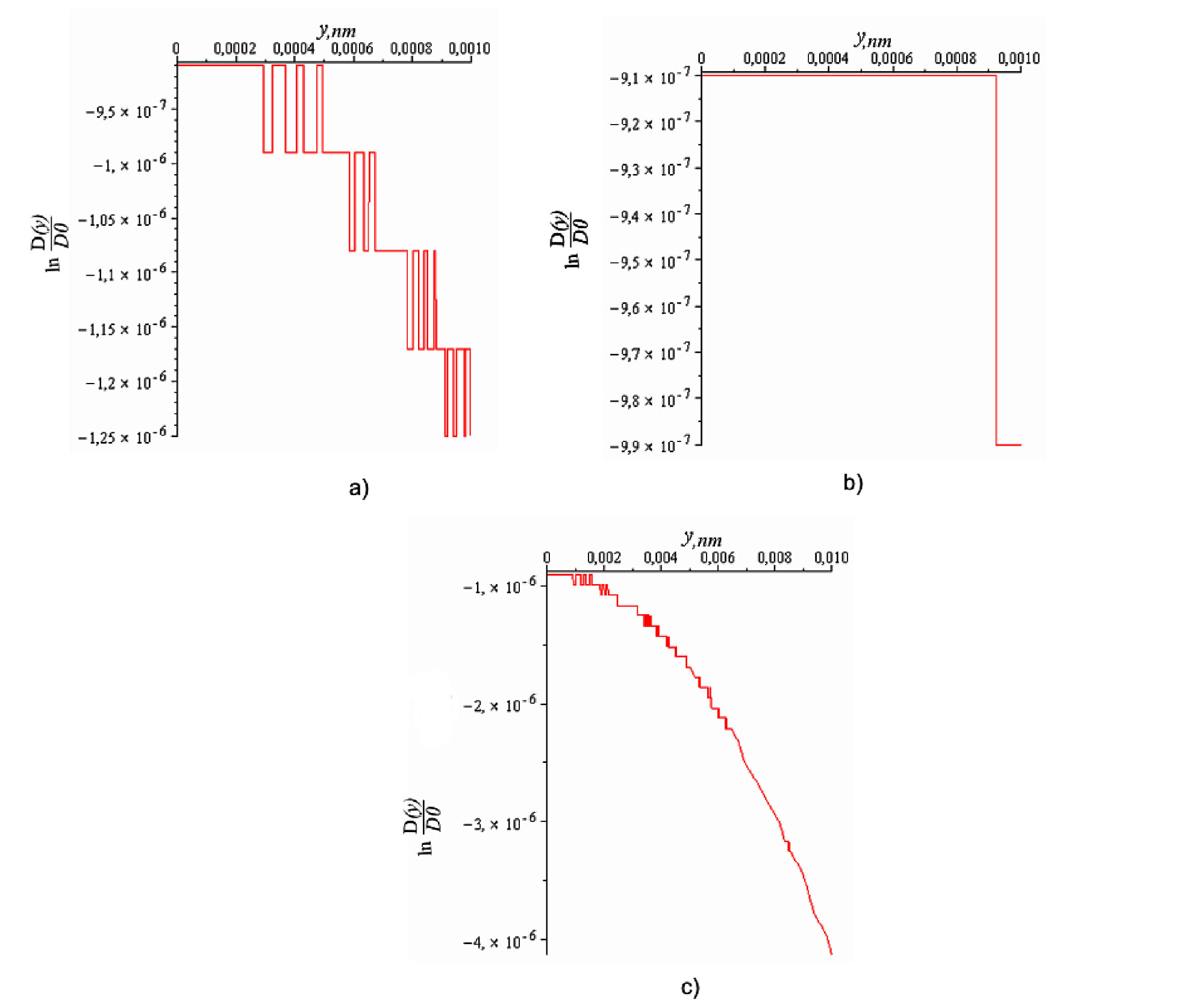}}
\caption{Dependence of the tunneling probability $\ln
\left[{D(y)}/{D_0}\right]$ on  $y$ for pores with ring~(a) and
elliptic~(b),~(c) profiles.}\label{fig-smp8}
\end{figure}
Calculations of the tunneling probability were carried out using
the program Maple-13 with the parameters, $U_{1} =40$~eV, $U_{2}
=1$~eV, $E=4$~eV, $R=10$~nm, $r=8$~nm, and semiaxis of the ellipse
$b^{2} =10$. The obtained results are shown in
figure~\ref{fig-smp8} in the form
$f(y)=\frac{D\left(y\right)}{D_{0} } $, where $\ln D_{0}
=51.320595$. Figures~\ref{fig-smp8}~(a),~(b), show $f(y)$ in both
models in the same range of $y$, and  in figure~\ref{fig-smp8}~(c), the dependence $f(y)$ in the model (b) in a wider range of
 $y$. In such  $y$ range in the model  {a)},
$f(y)$ has a similar form but with less marked nonmonotonicities.
Due to a large range change of $f(y)$, its real form is somewhat
garbled by the applied Maple-13 spline. Nevertheless, even this
form allows us to observe the manifestations of resonant tunneling
in the models.

\section{Discussion of results. Conclusions}

The obtained dependencies  {$\ln \frac{D(y)}{D_0}=f (y)$} (see
figure~\ref{fig-smp8}) are generally of decreasing character with
a set of repeating plateaus.

The decrease of tunneling probability with an $y$ increase is
apparent.  For any fixed $y$ it corresponds to the one-dimensional
eigenvalue problem with two identical square barriers. The
increase of parameter $y$ in our case causes the widening of the
barriers and at the same time the narrowing of the hollow. It is
known that in case of tunneling through a single barrier, the
value of tunneling, similarly to our problem, decreases but this
dependence does not have  plateaus, which we relate to the
resonant tunneling.

Let us consider the virtual states of an electron in a hollow,
i.e., between two barriers. In order to qualitatively comprehend
the obtained behavior of the tunneling probability, it is
sufficient to employ the results of a quantum-mechanical problem
of a particle in infinitely deep square well. It is well known
that here the energy spectrum of a particle is a set of discrete
levels $E_{n} =\frac{\pi ^{2} \hbar ^{2} }{2md^{2} } n^{2} $ ($m$
is the mass of a particle, $n$ is the principal quantum number,
$n=1,\, 2,\, 3,\dots $). In our case, the dependence of the well
width $d$ on $y$ (and consequently $E_{n} $), causes the observed
behavior of the tunneling probability. Compare the behavior of the
ground states $E_{1} \left(y\right)$ for two types of the pore
cross section, i.e. the circular  and more general elliptic one
extended along $OY$. In the first case, $E_{1a}
\left(y\right)=\frac{\pi ^{2} \hbar ^{2} }{2m\left(r^{2} -y^{2}
\right)} $,  and in the second case $E_{1b}
\left(y\right)=\frac{\pi ^{2} \hbar ^{2} }{2m\left[r^{2} -({y}/{b}
)^{2} \right]} $, ($b\geqslant 1$ is the semi-axis of an
ellipse). It can be seen that upon the axis, i.e., at $y=0$, these
levels coincide. However, the change $y\to y+\delta y$ causes a
more rapid rise of position of the level $E_{1a}(y)$ in comparison
with the growth of $E_{1b}(y)$. Within the limit $b\to \infty $,
$E_{1b} (y)$ does not depend on $y$, i.e., this will be a level it
``degenerates'' by $y$. In a general case, a step-by-step
transition $y\to y+\delta y$ is accompanied by the appearance of
closely positioned levels.

As noted above, at $y=0$,  the values $f(y)$  coincide for both
models. In the present case, $f(y)$ is nonzero. It means that the
energy of the incident electron $E$ differs from the virtual
levels. Suppose that $E$ is between the virtual levels $E_{n} $
and $E_{n+1}$. Increasing  {y }(which is accompanied by
narrowing the cavity) the virtual levels rise on the energy scale.
In model {a)} the $n$-th level  $E(y)$ with $y\simeq 0.00029$
coincides with energy $E$ of the incident electron. As a result,
there is a sharp drop $f(y)$, i.e. the resonant tunneling effect.
A similar behavior of $f(y)$ is observed  in the case  { b)}, but
for $E(y)$ at $y\simeq 0.00093$. The difference between these
values
 $y$ is connected with~the fact that the hollow thickness in
the model   {a}) changes more rapidly with the change of
 $y$ than in the model  {b)}. A further increase of
 $y$, for example, in the model  {a)} is accompanied
by resonant tunneling due to the levels $E_{n} $ at $E(y)$ for
$y$ up to its value $\sim 0.0006$. At $y\simeq$0.0006, the next
virtual level, $E_{n-1} $, coincides with energy  $E$  of the
incident electron. Further,  the non-monotonous series of
$f(y)$follows up to $y\simeq 0.0008$ at which the next virtual
level,  $E_{n-2} $, coincides with $E$ and so on.

The main conclusion of the paper consists in stating that the
degree of the electron tunneling into the hollow is determined
both by the geometry of the pore and by the energy of the
bombarding electron.

The obtained results may be represented alternatively. Let us
consider a homogeneous flow parallel to the axis $y=0$ of
monochromatic electrons. In the range of $y$ where virtual states
coincide or are close to the energy of bombarding electrons, their
contribution in tunneling into hollow is  either altogether absent
or just minimal. A more real case is the flow of electrons with
some distribution by energy. Assume that this is Gaussian-type
distribution with a maximum at $E=E_{\max }$\,. Even if $E_{\max }$
coincides with virtual level, there are the energies from the
``tails''  of the Gaussian that do not coincide with this level.
Exactly these electrons ensure the tunneling. However, since the
number of such electrons according to the Gaussian distribution is
low, their contribution into tunneling will be negligible. In
other words, a more real picture being taken into account does not
change the qualitative conclusions obtained earlier.

Herein above we have focused  on the maximum tunneling taking
place at $y=0$ in case of a single barrier. The presence of the
second identical potential barrier and hence a possible resonant
tunneling, somewhat changes this statement. In the case of
geometric parameters of the pore, such that at $y=0$, its virtual
levels coincide with the energy of the electron, this electron
does not provide any contribution into the tunneling. The
tunneling is realized by electrons that are more distant from
$y=0$ and their energies do not coincide with the virtual levels.
Thus, due to a larger thickness of barriers in comparison with the
$y=0$ case, the degree of  tunneling will be much lower. However,
suffice it to change the energy of the electrons so that it does
not coincide with the values of virtual states in the region
$y=0$, then the degree of tunneling will abruptly grow. On the
other hand, the degree of tunneling may be essentially different
for the electrons with the same energy depending on the geometry
of the pores. In other words, the degree of tunneling is not much
effected by the energy of electrons or by the geometry of pores,
but rather by their interrelation. Thus, the above mentioned
result of the paper~\cite{7} concerning the maximum tunneling in
9-nm pores in comparison with the 5, 11, the 14-nm pores may be
caused by the energy bombarding electrons which least of all
coincides with their virtual states.

\begin{figure}[ht]
\centerline{\includegraphics[width=0.8\textwidth]{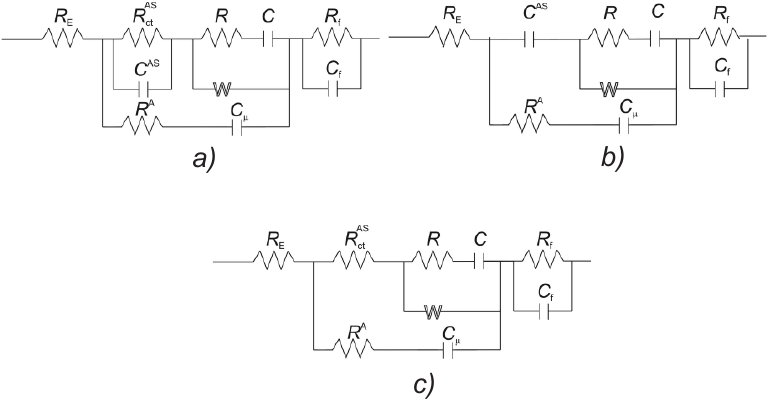}}
\caption{Equivalent electric circuits of impedance model for the
Li$^{+}$-intercalation in the nanodispersed silicon dioxide: a~--
common, b~-- at the interference blockade of electron tunneling,
c~-- in presence of electron tunneling.}\label{fig-smp9}
\end{figure}

The obtained results make it possible to propose the impedance
model of the investigated mechanisms presented in
figure~\ref{fig-smp9}~(a). Here  {$R{}_{\mathrm{E}}$} is the resistance of
electrolyte,  {$R{}^{A}$ } is a relaxation resistance of the
low-resistance phase  ${A }$ and $C_{\mu } $ is its chemical
capacity (chemical capacity is the ratio of carrier concentration
 {n} and Fermi level $E_{{\rm F}n} $, $C_{\mu } =e^{2}
\frac{\partial n}{\partial E_{{\rm F}n} }$~\cite{14}). Due to
quantization of an energy spectrum in nanoparticles, the chemical
capacity is transformed~into~the quantum~capacity~\cite{12}, and
it noticeably effects the transport processes~\cite{15}. This
effect should be expected only in nanoscale objects because in
this case only the quantum capacity can disconnect the circuit
R3-C3 (figure~\ref{fig-smp7}~(b)), i.e., the DC-mode.

The branched  $R_{ct}^{AS }\parallel C^{AS}$ circuit produces a
charge transport through the barrier
 $a$ and $R_{f }\parallel C_{f}$~-- through a
barrier $c$, which  divides silicon dioxide and electrolyte. The
Rendls-Ershler circuit~\cite{16} produces the lithium cations
transport from electrolyte into the structural channels
 {SiO${}_{2}$}.

\begin{figure}[h]
\centerline{\includegraphics[width=0.5\textwidth]{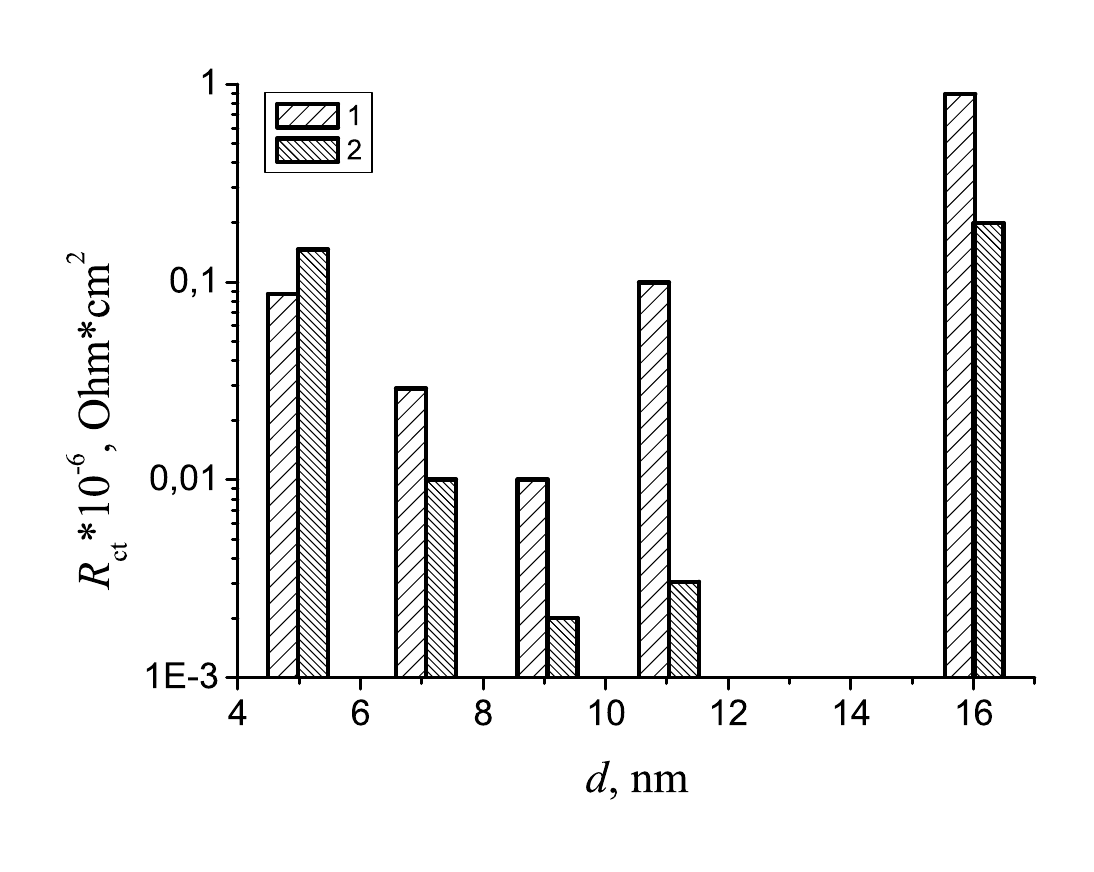}}
\caption{Dependence of the parameter
$R_{ct}$ on the particle size.}\label{fig-smp10}
\end{figure}

According to the results of theoretical calculations it is obvious
that  $R_{ct}^{AS}$ (as well as $C^{AS} $) depends on the particle
size.  Therefore, at maximum and minimum value
 $R_{ct}^{AS}$, the equivalent circuit 9~(a)
transforms into the circuits 9~(b) and 9~(c) (see
figure~\ref{fig-smp9}). It is easy to see that the circuit 9~(b)
(figure~\ref{fig-smp9}) is capacitive. It well describes the
impedance behavior at the electron tunneling blockade into the
nanoparticle  {SiO${}_{2}$} (figure~\ref{fig-smp4}, curve~1). The
circuit~9c corresponds to deblocking the electron tunneling and to
the formation of two-phase state. Results of the proposed model
well agree with a package of experimental data: Kramers-Kronig
test does not exceed $3\cdot10^{-5}$, and the frequency
dependencies of ordinary difference are of  completely random
character. According to the computer parametric identification
(figure~\ref{fig-smp10}), the difference between experimental and
model curves does not exceed~5\%.

Figure~\ref{fig-smp10} shows an essential dependence of the
parameter  {$R{}_{ct}$} on the particle size. Its sharp minimum at
$d = 9$~nm coincides with the anomaly $\frac{\rd x}{\rd U_{\rm
rel} } $ at this value. Here is another interesting fact.
Intercalated lithium into nanoparticles of various sizes
(excluding very small ones) increases the charge transport.

It should be noted that the absence of diffusion control in this
case and, as a consequence, the minimum content of the diffusion
resistance in the total resistance of a system, is extremely
important  to practical application of the 9-nm particles in the
current-generating process
\[
 {\rm SiO}_{2} + x{\rm Li}^{+} + xe^{-} = {\rm Li}_{x}{\rm SiO}_{2}
\]
with a horizontal discharge curve at $0.9<x<2$. By contrast,
another situation for the 5-nm particles (obviously, for the 16-nm
particles as well) is observed~-- the positive charged lithium is
screened by electrons, injected from contact to a conductive
agent. Such an effect (obviously, possible only for nanoscale
objects) may be defined as sub-Faraday or pseudo-Faraday discharge
with the expected hypercapacity. Actually, at discharge of a
galvanic cell with electrodes on the base of both the 5-nm and
16-nm silicon dioxide, the horizontal plots in voltage-time
characteristics are absent. It testifies to the process without
the 1-st type of phase transition. Results of the potentiometric
measurements do not allow us to claim the prevalence of the
diffusion control over the kinetic one. At least in the range of
2.94--1.85~V, the dynamic current-voltage characteristic is close
to ideal polarizability. The calculated specific capacity of  ``a
condenser regime'' in a given electric voltage range for
nanocomposites on the basis of the 5~nm silicon dioxide was
$\sim$1075~F/g, and for the 16~nm one  it was 2034~F/g. Taking
into account the active surface area, differential capacitance
reaches the record value $\sim$256 and 2400 $\mu$F/cm$^{2}$,
accordingly. The latter value exceeds the known maximum value of
the differential pseudo-capacitance by about five times~\cite{17}.
As in the ideal polarizability case, the multifold charge-discharge
cycles are possible here.  This emphasizes the importance of the
above considered nanomaterials to the technology of
electrochemical storage with the high cycling. A similar
conclusion is less obvious for the 11-nm nanosize titanium
dioxide. Obviously, faraday and pseudo-Faraday processes co-exist
here.

\section{Conclusions}

\begin{enumerate}
\item
The nanosize silicon dioxide with the Li${}^{+}$-intercalated
current-generating reactions in it is suitable for technology of
electrical storage with the essentially higher specific capacities
and energy than in the traditional lithium current sources.
\item Processes in the materials depending on size of nanoparticles
can be accompanied by the formation or a two-phase state, or a
pseudo-Faraday process with great differential capacitance. The
reason for such a process is quantum-mechanical effect of
interference blockade of the electronic tunneling into the
nonmetallic nanoparticle. However, a mixed mechanism of
current-generating is not excluded.
\item Minimization of diffusion resistance in the nanosize silicon
dioxide particles promotes the achievement of high capacities.
\item The phenomenon of current-generating process in nanostructured
systems is analyzed on the basis of quantum mechanical tunneling.
The geometric and energy parameters, that simulate the
nanostructure, entail the different structure of the electron
virtual states. The efficiency of tunneling nontrivially depends
on the relation between these states and the energy of electrons
bombarding the nanostructure. Thus, the tunneling efficiency is a
non-monotonous function on the electron energy.
\end{enumerate}

\ukrainianpart

\title {╩трэЄютю-ьхїрэ│ўэх  тш∙х Єєэхы■трээ  │ хЇхъЄштэ│ёЄ№ ЇрЁрфх║тё№ъюую ёЄЁєьюєЄтюЁхээ  т  яюЁшёЄшї эрэюёЄЁєъЄєЁрї}
\author{┴.└. ╦єъ│ эхЎ№, ─.┬. ╠рЄєыър, ▓.▓. ├ЁшуюЁўръ}
\address{═рЎ│юэры№эшщ єэ│тхЁёшЄхЄ ``╦№т│тё№ър яюы│Єхїэ│ър'', тєы.~╤.~┴рэфхЁш,~12, 79013~╦№т│т, ╙ъЁр┐эр}

\makeukrtitle

\begin{abstract}
\tolerance=3000%
┬ ЁюсюЄ│ фюёы│фцхэю ЄхЁьюфшэрь│ўэ│ Єр ъ│эхЄшўэ│ чръюэюь│ЁэюёЄ│ яЁюЎхёє ы│Є│║тю┐ │эЄхЁъры Ў│┐ C--SiO${}_{2}$ эрэюъюьяючшЄ│т. ┬ёЄрэютыхэр чрыхцэ│ёЄ№  ъ фы  фшЇхЁхэЎ│щэю┐ ║ьэюёЄ│, Єръ │ фы  ъ│эхЄшўэшї ярЁрьхЄЁ│т яЁюЎхёє т│ф Ёючь│Ё│т эрэюъюьяючшЄ│т. ╟у│фэю ч Ёхчєы№ЄрЄрьш ЄхюЁхЄшўэюую рэры│чє, юЄЁшьрэ│ чрыхцэюёЄ│ ║ эрёы│фъюь ътрэЄютю-ьхїрэ│ўэюую хЇхъЄє │эЄхЁЇхЁхэЎ│щэю┐ сыюърфш хыхъЄ\-Ёюээюую Єєэхы■трээ  т эхьхЄры│ўэє эрэюўрёЄшэъє. ═ртхфхэр │ьяхфрэёэр ьюфхы№ яЁюЎхё│т Єр ярЁрьхЄЁшўэр │фхэЄшЇ│ърЎ│ . ╧ЁхфёЄртыхэр эютр Єхїэюыюу│  хыхъЄЁюї│ь│ўэшї ухэхЁрЄюЁ│т хэхЁу│┐.
\keywords эрэююс'║ъЄ, хыхъЄЁюээшщ ёЄрэ, Єєэхы■трээ , │эЄхЁъры Ў│ 

\end{abstract}


\begin{thebibliography}{17}
\bibitem{1}Shao-Horn Y., Osmialowski S., Horn Q.C.,
J.~Electrochem. Soc., 2002., {\bf 149}, No.~11., A1499--A1502; \\
\doi{10.1149/1.1513558}.

\bibitem{2} Quintin M., Devos O., Delville M.H.,
Electrochim. Acta., 2006, {\bf 51}, 6426--6434; \\
\doi{10.1016/j.electacta.2006.04.027}.

\bibitem{3}  Brandt K. A critical review of rechargeable lithium
battery technology. -- In: Proc. 12-th Intern. seminar on primary
and secondary battery technology and application., Deerfield
Beach, USA, 1995.

\bibitem{4}Wells A.F., Structural Inorganic Chemistry.
5-th ed. Oxford University Press, Oxford, 1984.

\bibitem{5}  Julien C.M., Mater. Sci. Eng. R, 2003, {\bf 40}, No.~2, 47--102;
\doi{10.1016/S0927-796X(02)00104-3}.

\bibitem{6}  Kwon C.W., Hwang S.J., Poquet A., Treuil N., Campet G., Portier J., Choy J.H.
-- In: New trends in intercalation compounds for energy storage. Series
Mathematics, Physics and Chemistry, 2002, {\bf 61}, 439--446.

\bibitem{7} Myronyuk I.F., Lobanov V.V., Ostafijchuk B. K, Mandzuk V.I. Grygorchak I. I., Yablon' L.S.,
Phys. Chem. Sol. State, 2001, {\bf 2}, No.~4, 653--660 (in
Ukrainian).

\bibitem{8} Venhryn B.Ya., Grygorchak I.I., Kulyk Yu.O.,
Opt. Appl., 2008, {\bf XXXVIII}, No.~1, 119--125.

\bibitem{9} Thompson A.G., Physica~B+C, 1980, {\bf 99B}, No.~4,  100--105;
\doi{10.1016/0378-4363(80)90216-8}.

\bibitem{10} Modern battery technology.  Ed. Tuck C.D.S.,
Ellis Horwook, New York,  1991.

\bibitem{11} Golant E.I., Pashkovskii A.B.,  Semiconductors, 2002, {\bf 36}, 311--318;
\doi{10.1134/1.1461409}.

\bibitem{12}Bohm D., Quantum Theory. Prentice Hall, New York, 1952.

\bibitem{13}Davies John H., The Physics of Low-Dimensional Semiconductors:
An Introduction. Cambridge University Press, 1998.

\bibitem{14} Mora-Sero I., Bisquert J.,
Nano Lett., 2006, {\bf6}, No.~4, 640--650;
\doi{10.1021/nl052295q}.

\bibitem{15}Luryi S., Appl. Phys. Lett., 1988., {\bf 52}, No.~8,
501--503; \doi{10.1063/1.99649}.

\bibitem{16} Stojnov Z.B. et al., An Electrochemical Impedance.
Nauka, Moscow, 1991 (in Russian).

\bibitem{17} Conway B.E., Electrochemical Supercapacitors. Plenum Publ., New York, 1999.
\end{thebibliography}
\end{document}